\documentclass[10pt,twocolumn,aps,prl,superscriptaddress,reprint]{revtex4-2}
\usepackage{graphicx}
\usepackage{float}
\usepackage{dcolumn}
\usepackage{hyperref}
\usepackage{multirow}
\usepackage{xcolor}
\usepackage{color}
\usepackage{soul,xcolor}
\usepackage{lineno}
\usepackage{amsmath}

\begin{document}

\title{Charge orders in fully intercalated bilayer TaSe$_2$:\\Dependence on interlayer stacking and intercalation sites}

\author{Yuhui Yan}
 \affiliation{College of Physics \& Optoelectronic Engineering, Department of Physics, Jinan University, Guangzhou 510632, China}

\author{Lingxiao Xiong}%
 \affiliation{College of Physics \& Optoelectronic Engineering, Department of Physics, Jinan University, Guangzhou 510632, China}

\author{Feipeng Zheng}%
 \email{fpzheng\_phy@email.jnu.edu.cn}
 \thanks{corresponding author}
 \affiliation{College of Physics \& Optoelectronic Engineering, Department of Physics, Jinan University, Guangzhou 510632, China}

\date{\today}

\begin{abstract}

Recent advancements have established self-intercalation as a powerful technique for manipulating quantum material properties, with precise controllable intercalation concentrations.
Given the inherently rich phase diagrams of transition metal dichalcogenides (TMDCs), studying the self-intercalated TMDCs can offer promising candidates for investigating the interplay between various orderings.
This work focuses on fully intercalated bilayer TaSe$_2$ (Ta$_3$Se$_4$), which has recently been fabricated experimentally.
By performing first-principles calculations, we demonstrate the suppression of  an intrinsic $3\times3$ charge density wave (CDW) in parent TaSe$_2$ layers, and emergence of $2\times 2$, $\sqrt{3} \times \sqrt{3}$, or absence of a CDW in the intercalated layers, depending on the stacking sequence and intercalation sites being occupied.
Particularly, the $2\times 2$ CDW shows an increase in electronic states at the Fermi level compared to its non-CDW phase.
This unusual behavior contrasts with that of typical CDW materials in TMDCs.
Furthermore,  superconductivity is preserved in these Ta$_3$Se$_4$ structures, with superconducting transition temperatures comparable to those of TaSe$_2$.
Spin-orbit coupling  is found to enhance the density of states at Fermi levels while simultaneously reducing the electron-phonon coupling  matrix elements. These two competing effects result in varying impacts on superconductivity across different Ta$_3$Se$_4$ structures.
Moreover, our calculations indicate that magnetic order is absent.
Our study deepens the understanding of underlying physics in Ta$_3$Se$_4$, and provides  experimentally feasible candidates  for studying  CDW, superconductivity, and their interplay.
\end{abstract}

\maketitle

\section{introduction}

Transition metal dichalcogenides (TMDCs) have attracted considerable attention due to their rich electronic phase diagrams. 
Recent studies suggest that adjusting interlayer coupling offers an effective approach to tailoring the properties of layered materials.
For example, by intercalating molecules into the interlayer space of bulk 2$H$-NbSe$_2$, it is possible to retain its high superconducting transition temperature while simultaneously exhibiting the Ising superconductivity characteristic of its monolayer counterpart \cite{zhang2022tailored,sun2023high}. 
Additionally, when alkali metal atoms are intercalated into the interlayers of transition metal compounds, these systems can transform from ordinary metals \cite{wu2021enhanced,zheng2020emergent,agarwal2023quasi,liu2021superconductivity,fan2019quick} or semiconductors \cite{hong2024multigap,woollam1977physics,huang2016dynamical} into superconductors, with relatively high superconducting transition temperatures ($T_{\mathrm{c}}$) or exhibiting Ising superconductivity. 
Furthermore, a transition from indirect to direct band gap semiconductor can be realized by eliminating the interlayer coupling in 2$H$-MoS$_2$ \cite{ellis2011indirect}.

Recently, the intercalation  of transition metal atoms can be realized with controllable intercalation concentrations by atomic diffusion in heterostructures~\cite{Wei2024}, ionic liquid gating~\cite{Wang2023}, or adjusting the relative chemical potential between metal and non-metal atoms~\cite{zhao2020engineering,pan2022,yang2024advancements,He2024}, which provides new insights for tuning material properties. 
When the intercalants belong to the internal component in TMDCs, these systems are referred to as self-intercalated TMDCs. 
The experimental and theoretical results indicate that full self-intercalation (100\%) significantly enhances superconductivity in PdTe$_2$~\cite{Wang2023,Wei2024}.
Magnetic ordering can be induced in heteroatoms-intercalated layered materials, as demonstrated by experimental results, particularly resulting in the anisotropy-stabilized long-range ferromagnetism in Fe-intercalated TaS$_2$ \cite{He2024}.
Furthermore, theoretical calculations suggest that self-intercalated 2$H$-TaS$_2$ systems can exhibit intercalation-concentration-dependent magnetic orders \cite{zhao2020engineering}, as well as  distinct  CDW orders from those of their parent materials \cite{hong2024multigap}, and can also switch between different structure phases when subjected to strains \cite{hong2024multigap}.
For 2$H$-TaSe$_2$, a sister compound of TaS$_2$, researchers have observed a $\sqrt{3} \times \sqrt{3}$ charge density modulation in a fully self-intercalated TaSe$_2$, using scanning transmission electron microscopy \cite{zhao2020engineering}. 
Considering that pristine TaSe$_2$ is a nonmagnetic system exhibiting coexistence of CDW and superconductivity, with $T_{\mathrm{c}}$s  and CDW transition temperatures ($T_{\mathrm{CDW}}$s) measured at approximately 0.1--0.15 K and 120--122 K \cite{wilson1975charge,bhoi2016interplay,freitas2016strong,PhysRevLett.34.734,neto2001charge,dai2014microscopic}, respectively, the aforementioned observation raises the following questions:
(1) Does the charge density modulation observed experimentally accompany a structure distortion? If so, what is the crystal structure of the distorted phase, and is the CDW enhanced or suppressed compared to pristine TaSe$_2$?
(2) How does self-intercalation affect superconductivity?
(3) Besides the experimentally observed structure, are there other competing structures for fully self-intercalated TaSe$_2$?
(4) Considering that Ta is a heavy element, what is the effect of spin-orbit coupling (SOC) in this system?

In this work, we will computationally show multiple charge orders in fully intercalated bilayer TaSe$_2$ (Ta$_3$Se$_4$). 
Ta$_3$Se$_4$ crystals can exhibit either the coexistence of $\sqrt{3} \times \sqrt{3}$, $2 \times 2$, and $2 \times 1$ CDWs with superconductivity, or superconductivity alone, depending on the interlayer stacking order  and the occupied intercalation sites in the interlayer space.  
In particular, our calculations accurately reproduce the experimentally observed Ta$_3$Se$_4$ crystal, which exhibits a specific crystal structure featuring a $\sqrt{3} \times \sqrt{3}$ CDW.  
We further show that this structure can host superconductivity with a $T_{\mathrm{c}}$ comparable to that of its parent material.
Furthermore, we find that the self-intercalation generally leads to the suppression of $3\times3$ CDW in parent TaSe$_2$, with an enhanced or an absent CDW in intercalation layers.
SOC is found to play diverse roles in the superconductivity of various Ta$_3$Se$_4$ crystals, exhibiting effects ranging from suppression and enhancement to negligible impact.
The underlying mechanisms behind these results are revealed through calculations of formation energies, electronic structures, harmonic and anharmonic phonons, electron-phonon coupling (EPC), and superconducting properties.

\section{Computational Methods}
{
First-principles density functional theory (DFT)  calculations were conducted using projector-augmented-wave (PAW)~\cite{PhysRevB.59.1758} and ultrasoft pseudopotential \cite{PhysRevB.41.7892}, combined with the PBEsol~\cite{PhysRevLett.100.136406} exchange-correlation functional.
}~These calculations employe a combination of the Quantum Espresso (QE)~\cite{RevModPhys.73.515,RevModPhys.89.015003,Giannozzi_2009} and Vienna Ab Initio Simulation Package (VASP)~\cite{PhysRevB.54.11169} software packages. 
The VASP package is used for calculating the unfolded band structures and Fermi surfaces, as well as  magnetic properties, while the remaining calculations are performed using the QE.
{
We focus on three crystal structures of Ta$_3$Se$_4$, named $1H_{\mathrm{hollow}}$-Ta$_3$Se$_4$, $2H_{\mathrm{Ta}}$-Ta$_3$Se$_4$, and $1H_{\mathrm{Ta}}$-Ta$_3$Se$_4$, respectively, which are defined later in the main text. 
The PAW pseudopotentials were used for the calculations of $1H_{\mathrm{hollow}}$-Ta$_3$Se$_4$ and $1H_{\mathrm{Ta}}$-Ta$_3$Se$_4$, while the ultrasoft pseudopotentials were used for $2H_{\mathrm{Ta}}$-Ta$_3$Se$_4$.
The reasons for the above choices can be found in Sec.~S1 \cite{SM}.
}
To simulate the thin-film geometry and reduce interactions between periodic boundaries, a vacuum layer of approximately 15~\AA~is introduced. 
The Kohn-Sham valence states are expanded using plane waves, with energy cutoffs set at 50 Ry and 500 Ry for wave functions and charge densities, respectively.
Structural optimizations are carried out until the Hellmann-Feynman force acting on each atom is less than 1 $\times$ $10^{-5}$ Ry/Bohr.
An $18\times18\times1$ $\boldsymbol{k}$-grid and a  $6\times6\times1$ $\boldsymbol{q}$-grid are used to calculate the ground states of charge densities and phonons, respectively, for non-CDW Ta$_3$Se$_4$ and Ta$_2$Se$_4$.
These grids are scaled for the calculation of CDW Ta$_3$Se$_4$ and Ta$_2$Se$_4$, according to the sizes of their respective CDW supercells.
The electron-phonon coupling (EPC) matrix elements $g_{mn, \nu}(\boldsymbol{k}, \boldsymbol{q})$ are first computed  \cite{PhysRevB.65.035109,PhysRevB.56.12847,MOSTOFI2008685,PONCE2016116} based on the above $\boldsymbol{k}$ and $\boldsymbol{q}$ grids,  where $m$ and $n$ are band indices, and $\nu$ indicates a phonon branch.
The matrix elements quantify the scattering amplitude between the electronic states with a wave vector $\boldsymbol{k}$, a band index $m$ [denoted as ($\boldsymbol{k},m)$], and ($\boldsymbol{k}$+$\boldsymbol{q}$, n)  through a phonon mode with a branch $\nu$ and a wave vector $\boldsymbol{q}$.
Then the matrix elements are further interpolated {\cite{MOSTOFI2008685} to a 180$\times$180$\times$1 $\boldsymbol{k}$-grid and a 60$\times$60$\times$1 $\boldsymbol{q}$-grid, whereby the Eliashberg function $\alpha^2F(\omega)$ is calculated.
The $\alpha^2F(\omega)$ is defined as
\begin{eqnarray}
\alpha^{2} F(\omega)=\frac{1}{2} \sum_{\nu} \int_{\mathrm{BZ}} \frac{\mathrm{d}\boldsymbol{q}}{\Omega_{\mathrm{BZ}}} \omega_{\boldsymbol{q}\nu} \lambda_{\boldsymbol{q}\nu} \delta\left(\omega-\omega_{\boldsymbol{q}\nu}\right),
\end{eqnarray}
where $\lambda_{\boldsymbol{q}\nu}$ is a phonon-momentum-resolved EPC constant, and $\Omega_{\mathrm{BZ}}$ is the volume of the first Brillouin Zone (BZ). 
The Dirac delta function $\delta\left(\omega-\omega_{\boldsymbol{q}\nu}\right)$ is approximated by a Gaussian function with a broadening of 0.5 meV.
The total EPC constants are obtained by
\begin{eqnarray}
\lambda = 2\int \frac{\alpha^{2} F(\omega)}{\omega}\mathrm{d}\omega.
\end{eqnarray}
This quantity can be effectively  represented as $\lambda = 2N(0)\langle |g|^2 \rangle /\omega_{0}$, where $N(0)$ is a density of states at Fermi level, and $\langle |g|^2 \rangle$ is an average of the EPC matrix elements on Fermi surface, which can be estimated by $\langle |g|^2 \rangle = \frac{1}{N(0)}\int\alpha^2\mathrm{F}(\omega)\mathrm{d}\omega$ \cite{PhysRevB.103.035411}.
The superconducting $T_{\mathrm{c}}$ is calculated by McMillan-Allen-Dynes  approach \cite{PhysRevB.12.905}, defined as
\begin{eqnarray}
    T_{\mathrm{c}}=\frac{\omega_{\mathrm{log}}}{1.2}\mathrm{exp}\left(-\frac{1.04(1+\lambda)}{\lambda-\mu^*(1+0.62\lambda)}\right),
\end{eqnarray}
where $\omega_{\mathrm{log}}$ is a logarithmic average of phonon frequencies $\omega_{\mathrm{log}}=\mathrm{exp}[\frac{2}{\lambda}\int_{0}^{\infty}\mathrm{d}\omega\frac{\alpha^2\mathrm{F}(\omega)}{\omega}~\mathrm{log}~\omega]$.
$\mu^*$ is the Morel-Anderson Coulomb potential, which is set to be 0.15, a typical value for TMDCs {\cite{luo2023emergent,PhysRevB.100.235420,Wu_2019}}.
{
  The temperature-dependent electron-momentum-resolved superconducting gaps on the Fermi surface, denoted as $\Delta(\boldsymbol{k},T)$, were determined by solving the anisotropic Migdal-Eliashberg equations on an imaginary axis and then analytically continuing to the real axis using Pad$\rm{\acute{e}}$ functions \cite{PhysRevB.87.024505}. 
  In solving these equations, the Kohn-Sham states within 100 meV of the Fermi level are included, and the Matsubara frequencies are cut off at 0.35 eV, which are sufficient to describe the EPC in this system.
}

Anharmonic phonon calculations are performed using the stochastic self-consistent harmonic approximation (SSCHA) \cite{PhysRevB.89.064302,PhysRevB.95.174104,monacelli2021stochastic}, a non-perturbative method that accounts for anharmonicity arising from both thermal and quantum fluctuations.
A $3\times 3$ supercell containing 63 atoms are chosen, which is commensurate with the $\boldsymbol{q}_{\mathrm{CDW}}$ of $1H_{\mathrm{hollow}}$-Ta$_3$Se$_4$.
We use a number of 2000 configurations in each population to obtain converged free energy Hessian.
Machine learning potentials, developed using the deep potential molecular dynamics method \cite{PhysRevLett.120.143001,WANG2018178}, are employed to model atomic interactions at various temperatures, enabling more efficient computation of the configurations in each population. 
The training sets comprise DFT-calculated energies, forces, and external pressures for one thousand configurations generated by SSCHA in each temperature.
The DFT calculations are performed  with the same precision as the electronic structure calculations described earlier.
To achieve optimal concordance between DFT-computed properties and those predicted by the machine learning potential, the loss function—encompassing energy, force, and external pressure contributions—undergoes minimization through four million iterative optimization steps.
A comparative analysis of energies, forces, and external pressures derived from DFT calculations and machine learning potential predictions shown in Sec.~S2 \cite{SM}.

The unfolded band structures and Fermi surfaces calculations are performed using the same method as our previous works \cite{zheng2018first,zheng2019electron}.

\section{results and discussions}

\subsection{Property of TaSe$_2$ without intercalation}

Before studying Ta$_3$Se$_4$, we first calculate the properties of TaSe$_2$ using various types of pseudopotentials to identify the most suitable computational methods for this system{, as shown in Sec.~S3 \cite{SM}}.
We find that using PAW pseudopotentials (\cite{PhysRevB.59.1758}) with an exchange-correlation functional of PBEsol  leads to the calculated hexagonal lattice constants $a=3.41$~\AA, and $c=12.63$~\AA, which closely match the experiments \cite{brown1965layer} with a maximum error of 0.32 \%.
The calculated electronic band structure for monolayer TaSe$_2$ displays a hole (electron) pocket centered at the $\Gamma$ (K) point, which is consistent with previous studies \cite{yan2015structural,ge2012effect,brown1965layer,si2019charge,lian2022intrinsic,lawan2023tuning}.
Furthermore, the calculated $\omega_{\boldsymbol{q}\nu}$ of bilayer TaSe$_2$, shown in Sec.~S4 \cite{SM}, exhibits the most negative phonon frequency at $\boldsymbol{q} = 2/3\mathrm{K}$, suggesting a $3\times 3$ CDW instability, consistent with experimental measurements \cite{yan2015structural,Wan2023,wilson1975charge}.
The above results suggest our computational method is valid to describe the properties in TaSe$_2$ systems.

\subsection{Three candidate crystal structures of Ta$_3$Se$_4$}

\begin{figure}[t]
    \centering
    \includegraphics[width=76 mm]{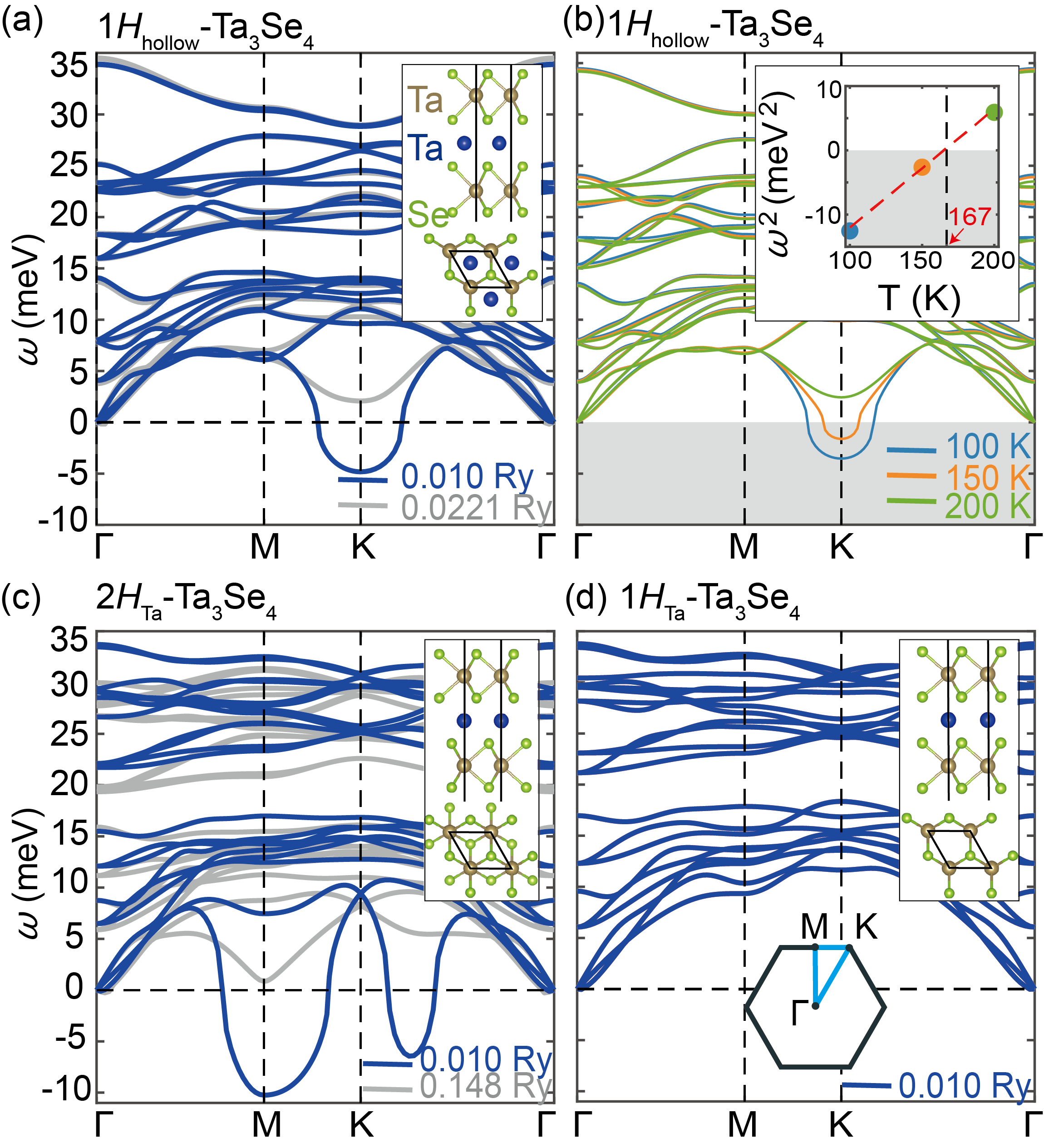}
    \caption{
    \label{fig1}
    Calculated phonon dispersions under harmonic approximation for (a) $1H_{\mathrm{hollow}}$-Ta$_3$Se$_4$, (c) $2H_{\mathrm{Ta}}$-Ta$_3$Se$_4$, and (d) $1H_{\mathrm{Ta}}$-Ta$_3$Se$_4$ in non-CDW phases, respectively, using different values of electron broadening. 
    The insets in the above panels show the side and top views of the corresponding structures. 
    (b) Anharmonic phonon spectra calculated from the SSCHA free energy Hessian at several temperatures for non-CDW $1H_{\mathrm{hollow}}$-Ta$_3$Se$_4$.
    The inset presents a linear fit of the squared phonon energies at the K point across different temperatures.
    }
\end{figure}

Following the intercalation of Ta, it is essential to determine the crystal structure of Ta$_3$Se$_4$, as the intercalant can potentially occupy multiple sites in the interlayer space, and the intercalation process may alter the stacking sequence between adjacent TaSe$_2$ layers.
We construct initial structural models that incorporate various stacking sequences {(Sec.~S5 \cite{SM})} and consider typical intercalation sites within the interlayer space.
After structure optimizations, we obtain {12} different structures, whose total energies are shown in Sec.~S6 \cite{SM}.
We found that two structures exhibit a tiny total energies difference of 13 meV/f.u., and their energies are substantially lower than those of the other structures.
Their calculated phonon dispersions ($\omega_{\boldsymbol{q}\nu}$s) and corresponding crystal structures are shown in Figs.~\ref{fig1}(c) and ~\ref{fig1}(d), respectively.
We refer to the former structure, shown in Fig.~\ref{fig1}(c), as $2H_{\mathrm{Ta}}$-Ta$_3$Se$_4$, since it features the $2H$ interlayer stacking between two TaSe$_2$ layers, and the occupied  intercalation sites aligned with the parent Ta atoms in the out-of-plane direction.
The latter one is referred to as $1H_{\mathrm{Ta}}$-Ta$_3$Se$_4$, since it manifests the same occupied intercalation sites but with $1H$ interlayer stacking (Fig.~\ref{fig1}(d)).
Upon further examination of the remaining structures, we found that one of them exhibits the same structural features as experimentally measured Ta$_3$Se$_4$ by scanning tunneling electronic microscopy \cite{zhao2020engineering}, but with a higher calculated energy of approximately 1.26 eV/f.u. related to $2H_{\mathrm{Ta}}$-Ta$_3$Se$_4$ or $1H_{\mathrm{Ta}}$-Ta$_3$Se$_4$. 
As shown in Fig.~\ref{fig1}(a), this structure is characterized by the $1H$ interlayer stacking and the occupied intercalation positions at the hollow sites, which we refer to as $1H_{\mathrm{hollow}}$-Ta$_3$Se$_4$.
Thus, although $1H_{\mathrm{hollow}}$-Ta$_3$Se$_4$ theoretically exhibits a higher total energy compared to both $1H_{\mathrm{Ta}}$-, and $2H_{\mathrm{Ta}}$-Ta$_3$Se$_4$, it can still be synthesized.
This is reminiscent of the synthesis of 2$M$-WS$_2$, which has a much higher energy than 2$H$-WS$_2$ \cite{paudyal2022superconducting,Fang_2020}.
The above results suggest the possibility of multiple structural phases in Ta$_3$Se$_4$, warranting further experimental validation.
Thus, the following discussions will focus on the three Ta$_3$Se$_4$ structures: $1H_{\mathrm{hollow}}$-, $1H_{\mathrm{Ta}}$-, and $2H_{\mathrm{Ta}}$-Ta$_3$Se$_4$.
The spin spiral calculations, as shown in Sec.~S7 \cite{SM}, suggest that all the three structures are non-magnetic, similar to the cases of fully intercalated TaS$_2$ in bilayer~\cite{zhao2020engineering,luo2023emergent} and bulk phases \cite{zhao2020engineering}.

\subsection{Property of $1H_{\mathrm{hollow}}$-Ta$_3$Se$_4$}

\begin{figure}[t]
    \centering
    \includegraphics{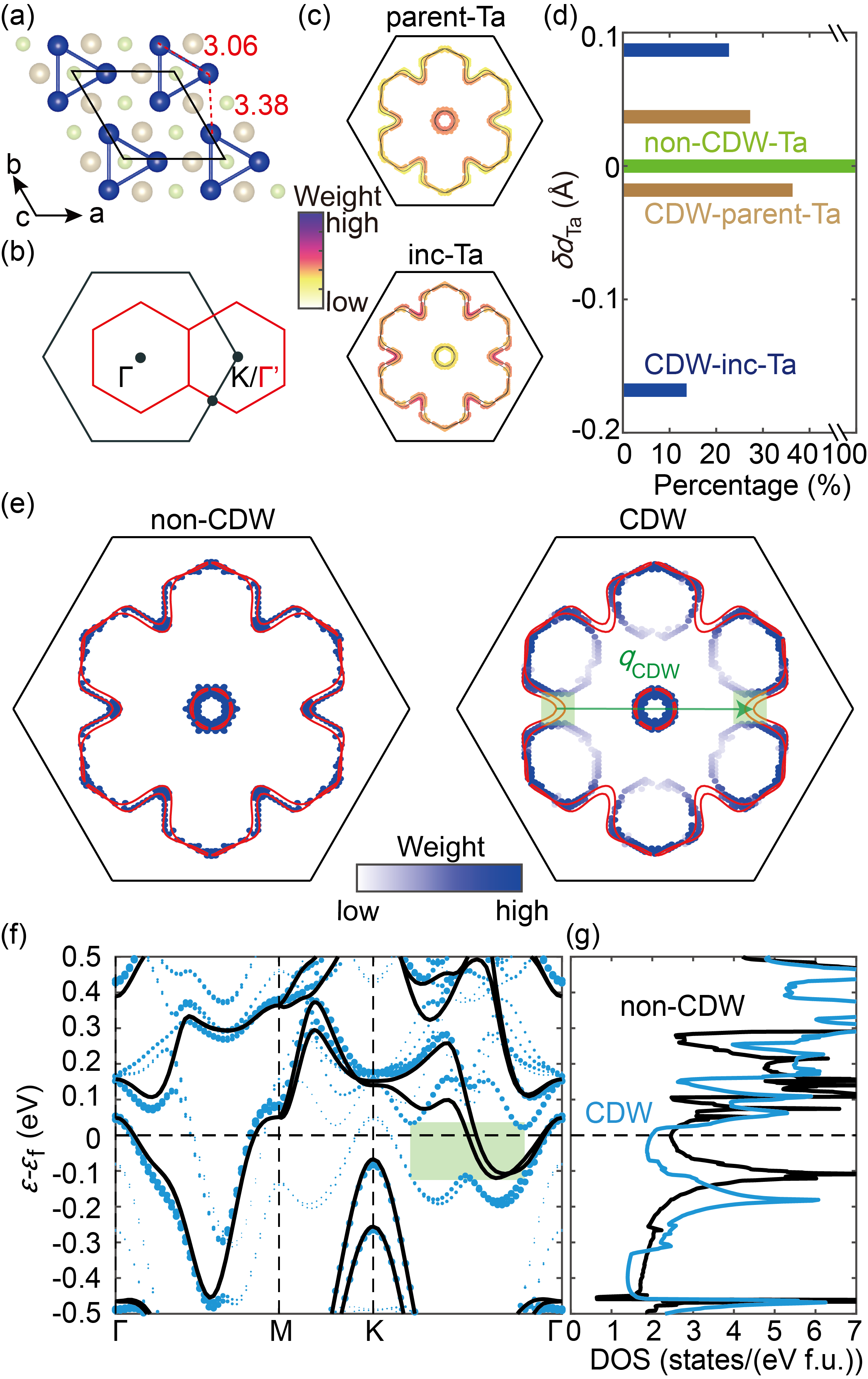}
    \caption{
    \label{fig2} 
    Crystal and electronic structure of $1H_{\mathrm{hollow}}$-Ta$_3$Se$_4$. 
    (a) Top view of its $\sqrt{3} \times \sqrt{3}$ CDW structure, where the intercalation layer is highlighted.
    The solid lines between Ta atoms in the intercalation  layer indicate that their distance decreases relative to that in non-CDW structure.
    More details can be found in Sec.~S8 \cite{SM}.
    (b) BZs for non-CDW (black) and CDW (red) phases, with corresponding high symmetry points. 
    (c) Projected electronic states around the Fermi surface onto Ta atoms in the parent TaSe$_2$ (parent-Ta) and intercalation  layer (inc-Ta). 
   (d) The distribution of $\delta d_{\mathrm{Ta}}$ in CDW $1H_{\mathrm{hollow}}$-Ta$_3$Se$_4$, showing the change of Ta-Ta distances compared to those in non-CDW phase that of its non-CDW phase. 
    (e) Unfolded Fermi surfaces of non-CDW and  CDW $1H_{\mathrm{hollow}}$-Ta$_3$Se$_4$ in primitive BZs.
    The green rectangular sections highlight two gapped regions that are related by $\boldsymbol{q}_{\mathrm{CDW}}$.
    Only the states within an energy window of $\pm$21 meV are shown.
    The solid lines represent the directly calculated Fermi surface of non-CDW structure.
    (f) The unfolded band structure of  CDW  $1H_{\mathrm{hollow}}$-Ta$_3$Se$_4$ in primitive BZ, overlaid with the directly calculated band structure of the non-CDW phase, represented by black lines.
    (g) DOS for non-CDW and CDW phases.
    }
\end{figure}

We begin with the analysis of lattice dynamics of $1H_{\mathrm{hollow}}$-Ta$_3$Se$_4$.
The calculated $\omega_{\boldsymbol{q}\nu}$, as shown in Fig.~\ref{fig1}(a), reveals the most negative phonon energy of -4.83 meV at the K point, indicating a potential $\sqrt{3}\times \sqrt{3}$ CDW instability in the undistorted $1H_{\mathrm{hollow}}$-Ta$_3$Se$_4$ crystal. 
This is in excellent agreement with the experimental measurement \cite{zhao2020engineering}, where the same CDW order has been detected.
Furthermore, we find that the frequency of the imaginary phonons is sensitive to the electronic broadening, indicating that they are associated with Kohn anomalies driven by EPC \cite{PhysRevLett.93.185503}.
As shown in Fig.~\ref{fig1}(a), slightly increase the electronic broadening from a regular value of 0.01 Ry to 0.022 Ry, eliminates those imaginary phonons.
This value of electronic broadening is slightly higher than that of bilayer TaSe$_2$ (0.021 Ry) as shown in Sec.~S4 \cite{SM}, implying a possible higher $T_{\mathrm{CDW}}$ in $1H_{\mathrm{hollow}}$-Ta$_3$Se$_4$.
To accurately determine $T_{\mathrm{CDW}}$,  it is essential to move beyond the harmonic approximation. 
This is achieved by considering both quantum ionic fluctuations and thermal anharmonic fluctuations at finite temperatures, utilizing SSCHA \cite{PhysRevB.89.064302,PhysRevB.95.174104,monacelli2021stochastic}.
The calculated  $\omega_{\boldsymbol{q}\nu}$s for $1H_{\mathrm{hollow}}$-Ta$_3$Se$_4$ at temperatures of 100, 150 and 200 K using SSCHA are shown in Fig.~\ref{fig1}(b).
As the temperature increases, the $\omega_{\boldsymbol{q}\nu}$s demonstrate a gradual hardening of the imaginary phonons around the K point, while the other phonon frequencies remain nearly unchanged.
By a linear fit of the $\omega_{\boldsymbol{q}\nu}^2$ at the K point, the $T_{\mathrm{CDW}}$ is estimated to be 167 K as shown in the inset of Fig.~\ref{fig1}(b).
The value of $T_{\mathrm{CDW}}$ is slightly higher than that of monolayer and bilayer TaSe$_2$, which are measured to be approximately 130~K~\cite{shi2018chemical,ryu2018persistent} and 120~K~\cite{wilson1975charge,bhoi2016interplay,PhysRevLett.34.734,neto2001charge}, respectively.
This is consistent with the relatively smaller CDW formation energy, and  more pronounced deformation of Ta-Ta distances in  CDW $1H_{\mathrm{hollow}}$-Ta$_3$Se$_4$, which will be discussed later. 

To further investigate the CDW structure associated with the imaginary phonon, we construct $\sqrt{3} \times \sqrt{3}$ supercells with random atomic displacements relative to the equilibrium positions.
After multiple structure optimizations, we obtain one structure with an energy of -4.95 meV/f.u. related to its non-CDW state.
This energy gain is larger than that of bilayer TaSe$_2$, calculated at -3.88 meV/f.u.,aligning with the calculated larger $T_{\mathrm{CDW}}$ in $1H_{\mathrm{hollow}}$-Ta$_3$Se$_4$ as mentioned before.
The calculated $\omega_{\boldsymbol{q}\nu}$ shown in Fig.~\ref{fig3}(c) indicates dynamical stability of the reconstructed structure, as all the imaginary phonons were removed after the atomic reconstruction.
This suggests that the reconstructed structure is likely to be a CDW structure induced by the phonon instability at the K point. 
By further examination of the $\sqrt{3} \times \sqrt{3}$ CDW structure, we find that the largest deviation of the Ta-Ta distances compared to that of its non-CDW  structure ($\delta d_{\mathrm{Ta}}$)  in the intercalation layer is -0.17~\AA, as shown in  Fig.~\ref{fig2}(d).
In contrast, the $\delta d_{\mathrm{Ta}}$ in parent layer reduces from 0.16~\AA~ to 0.04~\AA~ after the intercalation.
This indicates that after the Ta intercalation, the $3\times 3$ CDW in parent TaSe$_2$ layers is suppressed, and the $\sqrt{3} \times \sqrt{3}$ CDW is emergent in the intercalated Ta layer, reminiscent of similar behavior in Ta$_3$S$_4$~\cite{luo2023emergent}.
The modification of the Ta-Ta distances after the intercalation results in the formation of trimers within the intercalation layer, which are centered at Se atoms from the top view, as shown in Fig.~\ref{fig2}(a). 

Furthermore, it is crucial to understand the effect of CDW on the electronic structure. 
The non-CDW $1H_{\mathrm{hollow}}$-Ta$_3$Se$_4$ exhibits a metallic band structure with two bands crossing Fermi level along $\Gamma$--$\mathrm{M}$ and $\Gamma$--$\mathrm{K}$ paths as shown by solid lines in Fig.~\ref{fig2}(f). 
This leads to the formation of two small, degenerate hole pockets and two large electron pockets centered at the $\Gamma$ point (Fig.~\ref{fig2}(e)).
These Fermi pockets are associated with the hybridization of electronic states from the parent and intercalated Ta atoms  with varying mixing ratios, as shown in Fig.~\ref{fig2}(c). 
The two inner pockets are primarily contributed by parent Ta atoms, with intercalated Ta atoms playing a secondary role.
The two outer pockets exhibit an opposite behavior.
By comparing the band structure of non-CDW $1H_{\mathrm{hollow}}$-Ta$_3$Se$_4$ to the unfolded band structure of CDW $1H_{\mathrm{hollow}}$-Ta$_3$Se$_4$ (blue dots), it is clear that  the most notable change near Fermi level is gaps opening with a size of approximately 0.13 meV along the $\Gamma$--$\mathrm{K}$ path, as highlighted in Fig.~\ref{fig2}(f).
This leads to the development of gaps around the Fermi surface near the wave vectors $1/2\mathrm{K}$ ($1/2\mathrm{K'}$), as demonstrated through a comparison of the Fermi surfaces: in the CDW phase, the unfolding weights near these wave vectors vanish (Fig.~\ref{fig2}(e)), whereas the non-CDW phase exhibits a uniform distribution of the weights with large values along the directly calculated Fermi surface (red solid lines).
This leads to the reduction of $N(0)$ from 2.46 /eV/f.u. in non-CDW phase to 1.98 /eV/f.u. in CDW phase.
The formation of the gaps near $1/2\mathrm{K}$ ($1/2\mathrm{K'}$) is reasonable, as the $\boldsymbol{k}$ states in different gapped regions are related by $\boldsymbol{q}_{\mathrm{CDW}}$, as indicated in Fig.~\ref{fig2}(e).
For example, the gapped regions highlighted in Fig.~\ref{fig2}(e) are connected by a wave vector of $2/3\boldsymbol{a}^{*} - 1/3\boldsymbol{b}^{*}$, which corresponds to one of the CDW vectors (see Sec.~S13 \cite{SM} for details). 
This observation is similar to other CDW materials in $H$-type TMDCs, where the CDWs lead to the partial gap opening at Fermi level and reduce N(0) to some extents \cite{si2019charge,wang2024polymorphic,zheng2018first,zheng2019electron,lian2022intrinsic,luo2023emergent}.
The above results demonstrate that the primary effect of CDW is to open energy gaps around Fermi surface near the wave vectors of the $1/2\mathrm{K}$ ($1/2\mathrm{K'}$) points, leading to the reduced $N(0)$.

\begin{figure}[t]
  \includegraphics{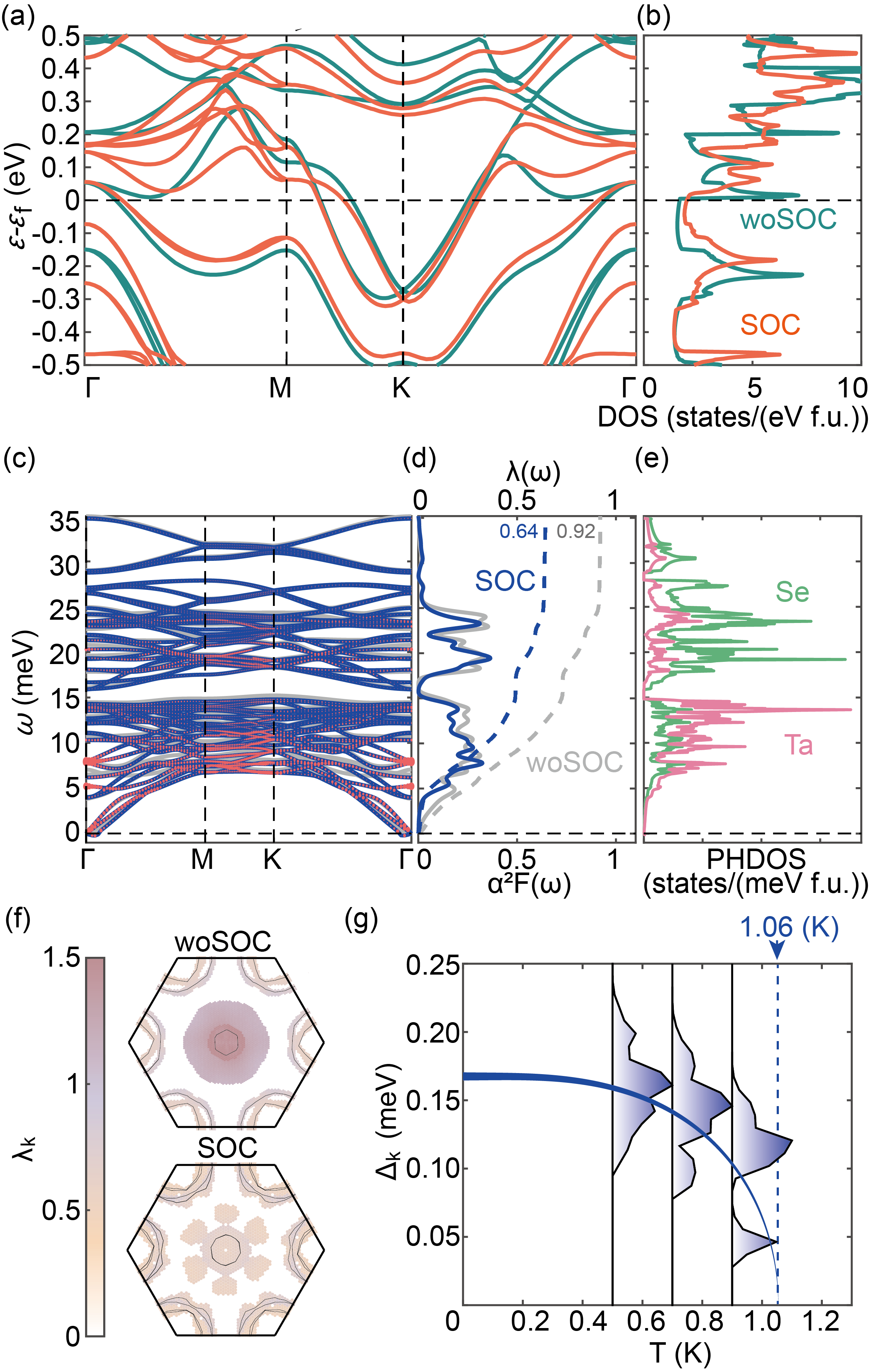}
  \caption{
  \label{fig3}
  The electronic properties and phonons of $1H_{\mathrm{hollow}}$-Ta$_3$Se$_4$ in the $\sqrt{3} \times \sqrt{3}$ CDW phase.
  The band structure (a) and DOS (b) of CDW $1H_{\mathrm{hollow}}$-Ta$_3$Se$_4$, calculated with and without SOC.
  (c) $\omega_{\boldsymbol{q}\nu}$ with (blue) and without (grey) SOC. 
  The size of red dots represents the value of $\lambda_{\boldsymbol{q}\nu}$ in the presence of SOC.  
  (d) $\alpha^2F(\omega)$ along with $\lambda(\omega)$, calculated with and without SOC.
  (e) Projected phonon density of states (PHDOS) onto the vibrations of Se and Ta atoms. 
  {
  (f) Distribution of $\boldsymbol{k}$-resolved EPC constants near the Fermi surface in the presence of SOC.
  Only the states within an energy window of $\pm$55 meV are shown.
  (g) Histograms of $\Delta(\boldsymbol{k},T)$ at various temperatures. 
  The blue curve represents a BCS fit of the energy gap.
  }
  }
\end{figure}

With the formation of gaps near the Fermi surface, a natural question arises: can superconductivity exist in  CDW  $1H_{\mathrm{hollow}}$-Ta$_3$Se$_4$?
As shown in Fig.~\ref{fig3}(d), the $\lambda$ of this system is calculated to be 0.64 with the inclusion of SOC.
Combined with the calculated $\omega_{\mathrm{log}} = 110.92$~K, the McMillan $T_{\mathrm{c}}$ ($T^{\text{ME}}_{\mathrm{c}}$) is estimated at 1.79 K.
{
Further consideration of the anisotropic EPC properties reveals a single gap structure of $\Delta (\boldsymbol{k},T)$, as shown in Fig.~\ref{fig3}(g).
The $\Delta (\boldsymbol{k},T)$ is predicted to vanish at 1.06 K, indicating the anisotropic Migdal-Eliashberg $T_{\mathrm{c}}$ ($T^{\text{aniso}}_{\mathrm{c}}$) around this value.
Furthermore, given that anharmonic effects may significantly impact the electron-phonon coupling \cite{Dangic2024}, we conducted further calculations on the electron-phonon coupling properties taking anharmonicity into account.
We found that both the phonon and $\alpha^2F(\omega)$ spectra remain nearly unchanged after accounting for anharmonicity, as shown in Sec.~S9~\cite{SM}.
This observation aligns with an experimental measurement indicating that the anharmonicity in layered TaSe$_2$ is weak \cite{Lin2020}.
}
According to previous studies, the $T_{\mathrm{c}}$ in TaSe$_2$ increases when going from bulk to monolayer limit within a range of 0.1--1.8 K \cite{wilson1975charge,bhoi2016interplay,freitas2016strong,lian2019coexistence,Wu_2019,galvis2013scanning,lian2023interplay}, as tabulated in Tab.~\ref{tab1}.
Thus, our result indicates that the superconductivity in $1H_{\mathrm{hollow}}$-Ta$_3$Se$_4$ can coexist with the $\sqrt{3} \times \sqrt{3}$ CDW order, with comparable $T_{\mathrm{c}}$ to its parent materials without intercalation. 

We also note that the SOC tends to suppress superconductivity in CDW $1H_{\mathrm{hollow}}$-Ta$_3$Se$_4$, {according to the results of both isotropic (Fig.~\ref{fig3}(d)) and anisotropic (Fig.~\ref{fig3}(g) and Sec.~S10 \cite{SM}) EPC calculations.}
{
For example, the calculated $\lambda$ with and without SOC are 0.64 and 0.92, respectively.
}
It is also worth noting  that SOC slightly increases $N(0)$ from 1.65 eV/f.u. to 1.98 eV/f.u., as shown in Fig.~\ref{fig3}(b).
Thus, the reduction in $\lambda$ is likely attributed to a decrease in the EPC matrix elements induced by SOC.
This is indeed the case, as incorporating SOC reduces the estimated average of the squared EPC matrix elements (see ``Computational Methods'')  from 2820 to 1850 meV$^2$. 
Furthermore, a comparison of the $\alpha^2F(\omega)$ between the two cases, as shown in Fig.~\ref{fig3}(d), clearly indicates that the reduction in  $\lambda$ originates from the diminished contribution of phonons in the 0--15 meV energy range.
These phonon states are related to the Ta vibrations as shown in Fig.~\ref{fig3}(e).
The above results indicated that the reduced EPC matrix elements are primarily associated with the vibrations of Ta atoms.

\subsection{Property of $2H_{\mathrm{Ta}}$-Ta$_3$Se$_4$}

The calculated $\omega_{\boldsymbol{q}\nu}$ for pristine $2H_{\mathrm{Ta}}$-Ta$_3$Se$_4$, as shown in Fig.~\ref{fig1}(c),  indicates that the most negative phonon energy occurs at the M point, with the second most negative energy appearing at the wave vector of $3/4\mathrm{K}$.
This suggests the competing CDW orders associated with the above two wave vectors.
Interestingly, an significantly large electron broadening of 0.148 Ry is needed to remove these imaginary phonons as shown in Fig.~\ref{fig1}(c), which is much larger than that of 2L-TaSe$_2$ (0.021 Ry) and  $1H_{\mathrm{hollow}}$-Ta$_3$Se$_4$ (0.022 Ry), suggesting a much higher $T_{\mathrm{CDW}}$ in $2H_{\mathrm{Ta}}$-Ta$_3$Se$_4$. 
This is similar to the case of $2H_{\mathrm{Ta}}$-Ta$_3$S$_4$, where a 0.08 Ry electron broadening is required \cite{luo2023emergent}.

\begin{figure}[h]
  \includegraphics[width=76 mm]{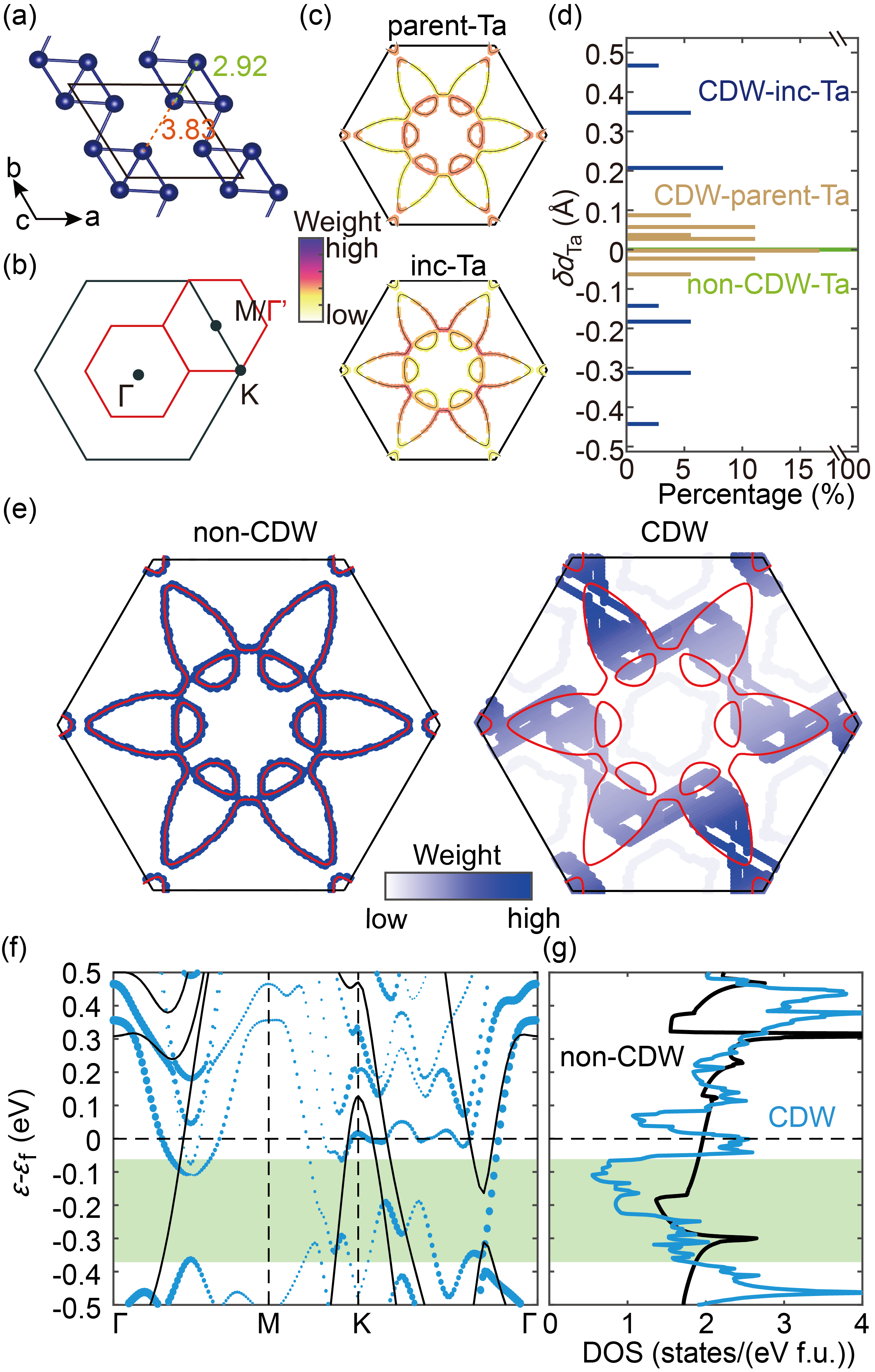}
  \caption{
  \label{fig4} 
  Crystal and electronic structure of $2H_{\mathrm{Ta}}$-Ta$_3$Se$_4$. 
  (a) Top view of the intercalation layer in $2 \times 2$ CDW structure. 
  More details can be found in Sec.~S8 \cite{SM}.
  (b) BZs for non-CDW (black) and CDW (red) phases. 
  (c) Projected electronic states around the Fermi surface onto parent-Ta and inc-Ta atoms. 
  (d) The distribution of $\delta d_{\mathrm{Ta}}$ in CDW $2H_{\mathrm{Ta}}$-Ta$_3$Se$_4$.
  (e) Unfolded Fermi surfaces of the non-CDW and CDW phases in primitive BZ.
  In panels (c) and (e), only the states within an energy window of $\pm$35 meV are shown.
  The solid lines are the directly calculated Fermi surface of the non-CDW phase. 
  (f) The unfolded band structure of the CDW phase  in primitive BZ, overlaid with the directly calculated band structure of the non-CDW phase, represented by black lines.
  (g) DOS for non-CDW and CDW phases.    
  The shaded areas in panels (f) and (g) highlight the energy ranges where band gaps emerge as a consequence of CDW formation.
  }
\end{figure}

As temperature decreases, the CDW transition is expected to initiate at the M points, due to the most negative phonon energy at these points.
Thus, we construct $2 \times 2$ supercells to study the corresponding CDW  distortion.
After multiple structure optimizations, we obtain a reconstructed structure with a total energy 47.45 meV/f.u. lower than non-CDW $2H_{\mathrm{Ta}}$-Ta$_3$Se$_4$.
This energy difference is an order of magnitude larger than that in the case of $1H_{\mathrm{hollow}}$-Ta$_3$Se$_4$.
The calculated $\omega_{\boldsymbol{q}\nu}$, shown in Fig.~\ref{fig5}(c), suggests the dynamical stability of this structure, as all $\omega_{\boldsymbol{q}\nu}$ values are positive, except for a small region near the $\Gamma$ point with negligible negative energies.
This suggests that the reconstructed structure is the CDW structure associated with the phonon instability at the M points.
The above result implies the CDW instability associated with $3/4\mathrm{K}$ is completely suppressed after the formation of the CDW relative to the M points.

Upon further analysis, the distribution of $\delta d_{\mathrm{Ta}}$  shows a behavior akin to that of $1H_{\mathrm{hollow}}$-Ta$_3$Se$_4$, as depicted in Fig.~\ref{fig4}(d): the CDW distortion is suppressed in parent TaSe$_2$, while enhanced in the intercalation layer,  with $\delta d_{\mathrm{Ta}}$ being less than 0.09~\AA~in the former and reaching 0.47~\AA~in the latter.
This $\delta_{\mathrm{Ta}}$  value is much larger than that of $1H_{\mathrm{hollow}}$-Ta$_3$Se$_4$ (0.17~\AA), consistent with the much larger $T_{\mathrm{CDW}}$ in  $2H_{\mathrm{Ta}}$-Ta$_3$Se$_4$, as previously analyzed.
The formation of the $\delta d_{\mathrm{Ta}}$ leads to a significant structure distortions relative to its non-CDW state in the intercalation layer, as shown in Fig.~\ref{fig4}(a).
The structure manifests quasi-one-dimensional Ta-Ta chains running along the direction of a lattice vector $\boldsymbol{b}$ as shown in Fig.~\ref{fig4}(a) (or its symmetry-equivalent directions).
Each of the chain consists of  edge-shared triangles  connected along the $\boldsymbol{a} + \boldsymbol{b}$ direction.
The above atomic reconstruction leads  to the formation of $2 \times 2$ CDW structure in $2H_{\mathrm{Ta}}$-Ta$_3$Se$_4$.
The main characteristic of the CDW $2H_{\mathrm{Ta}}$-Ta$_3$Se$_4$ is similar to $2H_{\mathrm{Ta}}$-Ta$_3$S$_4$, both featuring quasi-one-dimensional Ta-Ta chains in the intercalation layer \cite{luo2023emergent}.
However, the chains in $2H_{\mathrm{Ta}}$-Ta$_3$S$_4$ are formed in $2 \times 1$ supercells in strain-free case.
In Sec.~S11 \cite{SM}, we show that compressive strains tend to gradually suppress CDW in $2H_{\mathrm{Ta}}$-Ta$_3$Se$_4$.
By applying a 2\% compressive strains, the system can transform from the $2 \times 2$ CDW state into a $2 \times 1$ CDW state.
Further increasing the strain to 5.4\% will eliminate the CDW instability, triggering a transition  to a non-CDW phase.
In contrast, tensile strains further stabilize and enhance the $2 \times 2$ CDW order.
The above results indicate a rich structure phase diagram in $2H_{\mathrm{Ta}}$-Ta$_3$Se$_4$.

Due to the significant CDW distortions discussed above, the electronic structure in the CDW phase undergoes substantial changes.
Interestingly, we found that the CDW increases the $N(0)$ in $2H_{\mathrm{Ta}}$-Ta$_3$Se$_4$, contrasting with the commonly observed phenomenon in TMDCs.
As shown in Fig.~\ref{fig4}(g), $N(0)$ exhibits a clear increase from 1.96 eV/f.u. in non-CDW $2H_{\mathrm{Ta}}$-Ta$_3$Se$_4$ to 2.46 eV/f.u. in its CDW phase. 
A further comparison of the DOS between the non-CDW and CDW phases reveals that the gap opening due to the CDW occurs primarily in the energy range of approximately  -0.36 to -0.06 eV, as highlighted in Fig.~\ref{fig4}(g).
This shifts the electronic states to energy levels on both sides adjacent to the gapped region, leading to the formation of the DOS peaks around the Fermi level (Fig.~\ref{fig4}(g)).
This is further rationalized by comparing the band structures of the non-CDW and CDW phases in primitive BZ as shown in Fig.~\ref{fig4}(f), where the unfolding weights for the electronic states of the CDW phase are clearly reduced in the gapped energy region, and enhanced near Fermi level.
A direct comparison of the number of electronic states near the Fermi level provides further support for the above conclusion, as shown in Fig.~\ref{fig4}(e), where an increased number of states near the Fermi level can be observed in the CDW phase.
We also note that the symmetry of the Fermi surface in the CDW phase is reduced to $C_2$ symmetry compared to that in the non-CDW phase, which can be attributed to the breaking of $C_3$ symmetry in the CDW phase.
A consistent result can also be obtained by comparing the band structures in supercell BZ, as shown in Sec.~S12 \cite{SM}, where the number of bands of the CDW phase is clearly reduced in the gapped energy region.
Thus, we have shown that the energy gap formation in occupied states increases $N(0)$ in CDW $2H_{\mathrm{Ta}}$-Ta$_3$Se$_4$.
This is different from typical CDW materials in TMDCs (e.g., NbSe$_2$, TaSe$_2$), where gaps form at the Fermi level, leading to the reduced $N(0)$ \cite{si2019charge,wang2024polymorphic,zheng2018first,zheng2019electron,lian2022intrinsic,luo2023emergent}.
The unusual behavior of CDW is reminiscent of monolayer VSe$_2$, where unconventional CDW  with the full gap in the unoccupied states were confirmed both experimentally and theoretically \cite{vanEfferen2021}.

\begin{figure}[t]
  \includegraphics[width=76 mm]{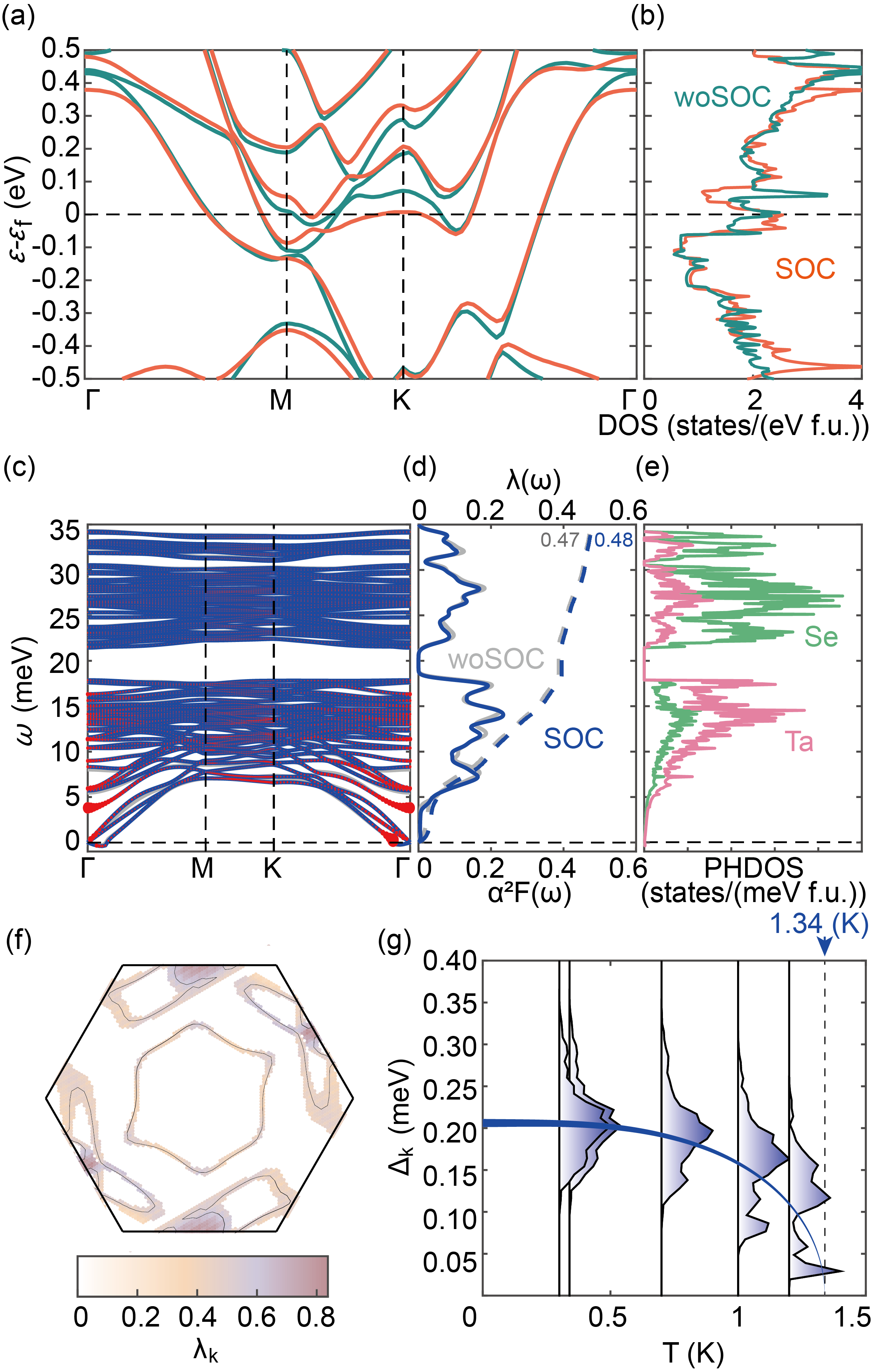}
  \caption{
  \label{fig5} 
  The  properties of $2H_{\mathrm{Ta}}$-Ta$_3$Se$_4$ in the $2 \times 2$ CDW phase.
  The band structure (a) and DOS (b) of the CDW phase, calculated with and without SOC.
  (c) $\omega_{\boldsymbol{q}\nu}$ with (blue) and without (grey) SOC. 
  The size of red dots represents the value of $\lambda_{\boldsymbol{q}\nu}$ in the presence of SOC.  
  (d) $\alpha^2F(\omega)$ along with $\lambda(\omega)$, calculated with and without SOC.
  (e) Projected PHDOS onto the vibrations of Se and Ta atoms. 
  {
  (f) Distribution of $\boldsymbol{k}$-resolved EPC constants near the Fermi surface in the absence of SOC.
  Only the states within an energy window of $\pm$35 meV are shown.
  (g) Histograms of $\Delta(\boldsymbol{k},T)$ at various temperatures in the absence of SOC.
  The blue curve represents a BCS fit of the energy gap.
  }
  }
\end{figure}

Although the CDW $2H_{\mathrm{Ta}}$-Ta$_3$Se$_4$ exhibits enhanced $N(0)$ compared to its non-CDW phase, a weak EPC in this material leads to a small $T^{\text{ME}}_{\mathrm{c}}$.
According to the calculated $\alpha^2F(\omega)$ and $\lambda(\omega)$, as shown in Fig.~\ref{fig5}(d), $\lambda$ of the system is only 0.48.
Combined with the calculated $\omega_{\mathrm{log}} = 109.06 $~K, the $T^{\text{ME}}_{\mathrm{c}}$ in CDW $2H_{\mathrm{Ta}}$-Ta$_3$Se$_4$ is estimated to be 0.39 K, which is clearly smaller than $1H_{\mathrm{hollow}}$-Ta$_3$Se$_4$ (1.79 K), and monolayer TaSe$_2$ (1.8 K), as tabulated in Tab.~\ref{tab1}.
Upon further comparison of $N(0)$ and $\lambda$ among the materials listed in Tab.~\ref{tab1}, we find that while the $N(0)$ of $2H_{\mathrm{Ta}}$-Ta$_3$Se$_4$ (2.46 /eV/f.u.) is comparable to  that of monolayer TaSe$_2$ (2.46 /eV/f.u.) and $1H_{\mathrm{hollow}}$-Ta$_3$Se$_4$ (1.98 /eV/f.u.), its $\lambda$ is noticeably smaller.
This can be attributed to the overall smaller EPC matrix elements in $2H_{\mathrm{Ta}}$-Ta$_3$Se$_4$, as its estimated $\langle |g|^2\rangle$ is only 1240 meV$^2$, which is substantially smaller than that of $1H_{\mathrm{hollow}}$-Ta$_3$Se$_4$ (1850 meV$^2$). 
We also find that SOC has minimal impact on the EPC and superconductivity in the CDW  $2H_{\mathrm{Ta}}$-Ta$_3$Se$_4$, reminiscent of similar behavior in CDW $2H_{\mathrm{Ta}}$-Ta$_3$S$_4$ \cite{luo2023emergent}.
The calculated $\alpha^2F(\omega)$ spectra are nearly identical with and without SOC (Fig.~\ref{fig5}(d)), and the values of $\lambda$ and $T^{\text{ME}}_{\mathrm{c}}$ show little variation with and without consideration of SOC, as listed in Tab.~\ref{tab1}. 
A further comparison of the values of $N(0)$ and $\langle |g|^2\rangle$ calculated with and without SOC (see Tab.~\ref{tab1}) reveals that the SOC slightly increases $N(0)$ and decreases $\langle |g|^2\rangle$.
These two effects counterbalance each other, resulting in similar values of  $\lambda$ and $T^{\text{ME}}_{\mathrm{c}}$.
We further show in Fig.~\ref{fig5}(a) that the enhancement of $N(0)$ induced by SOC is directly related to the band splitting near the Fermi level, which results in the formation of a locally flat band in the vicinity of the K point.

{
After the consideration of anisotropic EPC properties without SOC, the  $T^{\text{aniso}}_{\mathrm{c}}$ is estimated at 1.34~K, as shown in Fig.~\ref{fig5}(g), which is higher than its $T^{\text{ME}}_{\mathrm{c}}$.
The observed enhancement can be attributed to the anisotropic distribution of $\lambda_{\boldsymbol{k}}$, which exhibits relatively large values near the M points, as depicted in Fig.~\ref{fig5}(f).
The above analysis highlights the critical role of anisotropic EPC in determining the superconducting properties of this system.
Furthermore, when SOC is introduced, the band structure shown in Fig.~\ref{fig5}(a) suggests a redistribution of $\lambda_{\boldsymbol{k}}$: the values around the M points are expected to decrease, while those near the K points are anticipated to increase. 
However, due to the counterbalancing effects of these changes, $T^{\text{aniso}}_{\mathrm{c}}$ is likely to remain largely unaffected.
}

\subsection{Property of $1H_{\mathrm{Ta}}$-Ta$_3$Se$_4$}

\begin{figure}
  \includegraphics[width=76 mm]{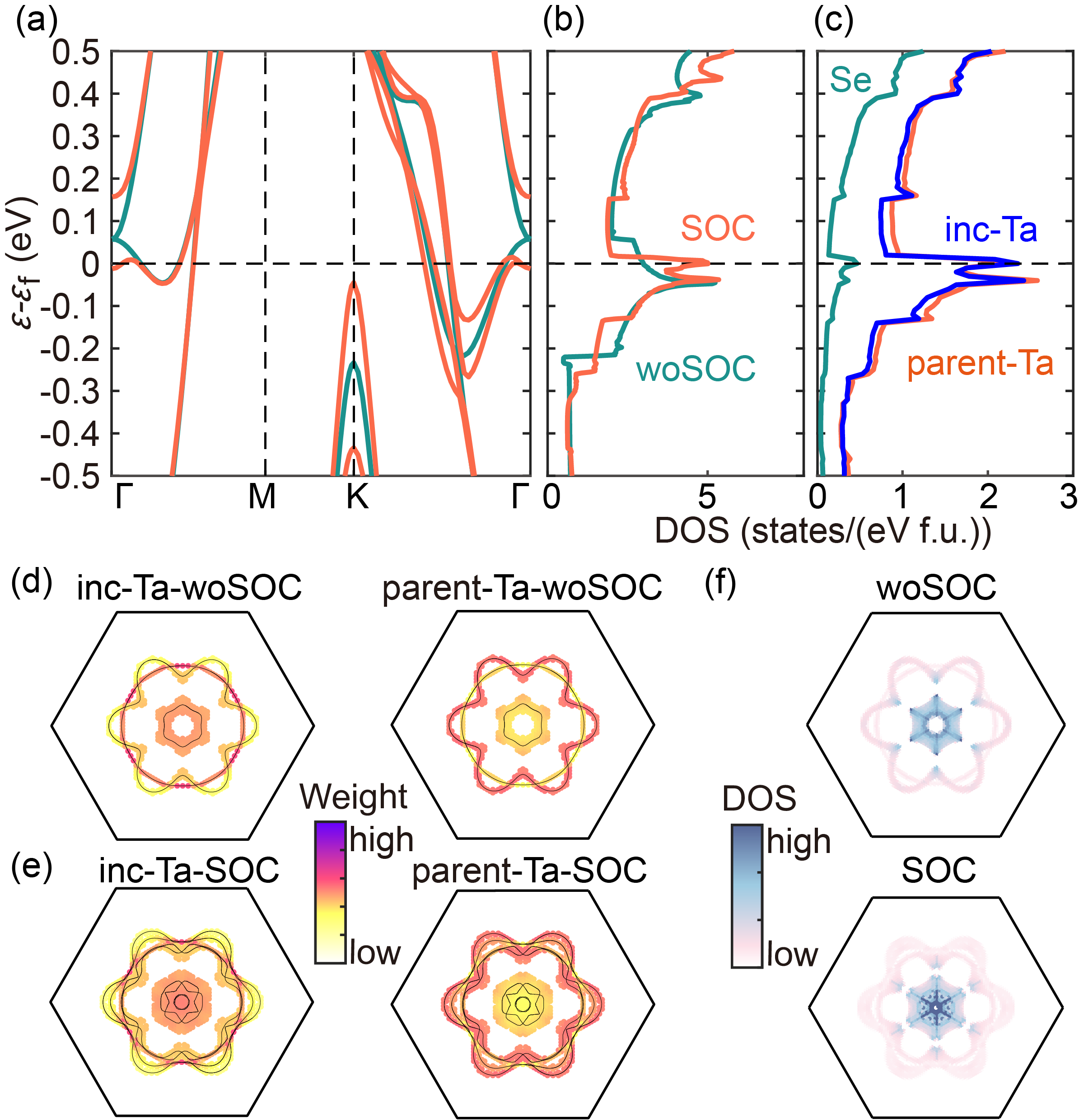}
  \caption{
  \label{fig6} 
  Electronic structure of $1H_{\mathrm{Ta}}$-Ta$_3$Se$_4$. 
  Band structures (a) and DOS (b), calculated with and without SOC. 
  (c) Projected DOS onto the electronic states of parent-Ta and inc-Ta atoms, calculated with SOC. 
  Projected electronic states around the Fermi surface onto parent-Ta and inc-Ta atoms, calculated without (d) and with (e) SOC. 
  Only the states within an energy window of $\pm$35 meV are shown.
  The solid lines represent the calculated Fermi surface.
  (f) Visualizations of the contributions  to $N$(0) from the electronic states in BZ, calculated with and without SOC.
  }
\end{figure}

Interestingly, the calculated $\omega_{\boldsymbol{q}\nu}$ for $1H_{\mathrm{Ta}}$-Ta$_3$Se$_4$ as shown in Fig.~\ref{fig1}(d) suggests its dynamically stable crystal structure, in contrast to the presence of CDW instabilities in $1H_{\mathrm{hollow}}$-Ta$_3$Se$_4$ and $2H_{\mathrm{Ta}}$-Ta$_3$Se$_4$.
This indicates the suppression of the $3 \times 3$ CDW in 1$H$-TaSe$_2$ following the intercalation of Ta atoms at the midpoint between each pair of  two parent Ta atoms that are vertically aligned in adjacent layers.

To further study the effect of the intercalation on the superconductivity, we tend to the electronic structure.
As shown in  Fig.~\ref{fig6}(a), 1$H_{\mathrm{Ta}}$-Ta$_3$Se$_4$ exhibits a metallic band structure, similar to that of $1H_{\mathrm{hollow}}$- and $2H_{\mathrm{Ta}}$-Ta$_3$Se$_4$.
There are several spin-degenerate bands crossing the Fermi level near the $\Gamma$ point when SOC is not included.
This leads to the formation of three Fermi pockets centered at the $\Gamma$: a small, hexagonal hole pocket, a circular electron pocket, and a flower-like electron pocket with six petals pointing to the BZ corners.
These  Fermi pockets are mainly contributed by the hybridization of Ta electrons with different mixing ratio, as shown in Fig.~\ref{fig6}(d).
The flower-like pocket is primarily contributed by the electronic states of the parent Ta atoms, with a relatively smaller contribution from the intercalated Ta.
In contrast, the other two Fermi pockets are mainly associated with the intercalated Ta, with the parent Ta playing a secondary role.

\begin{figure}[t]
  \includegraphics[width=76 mm]{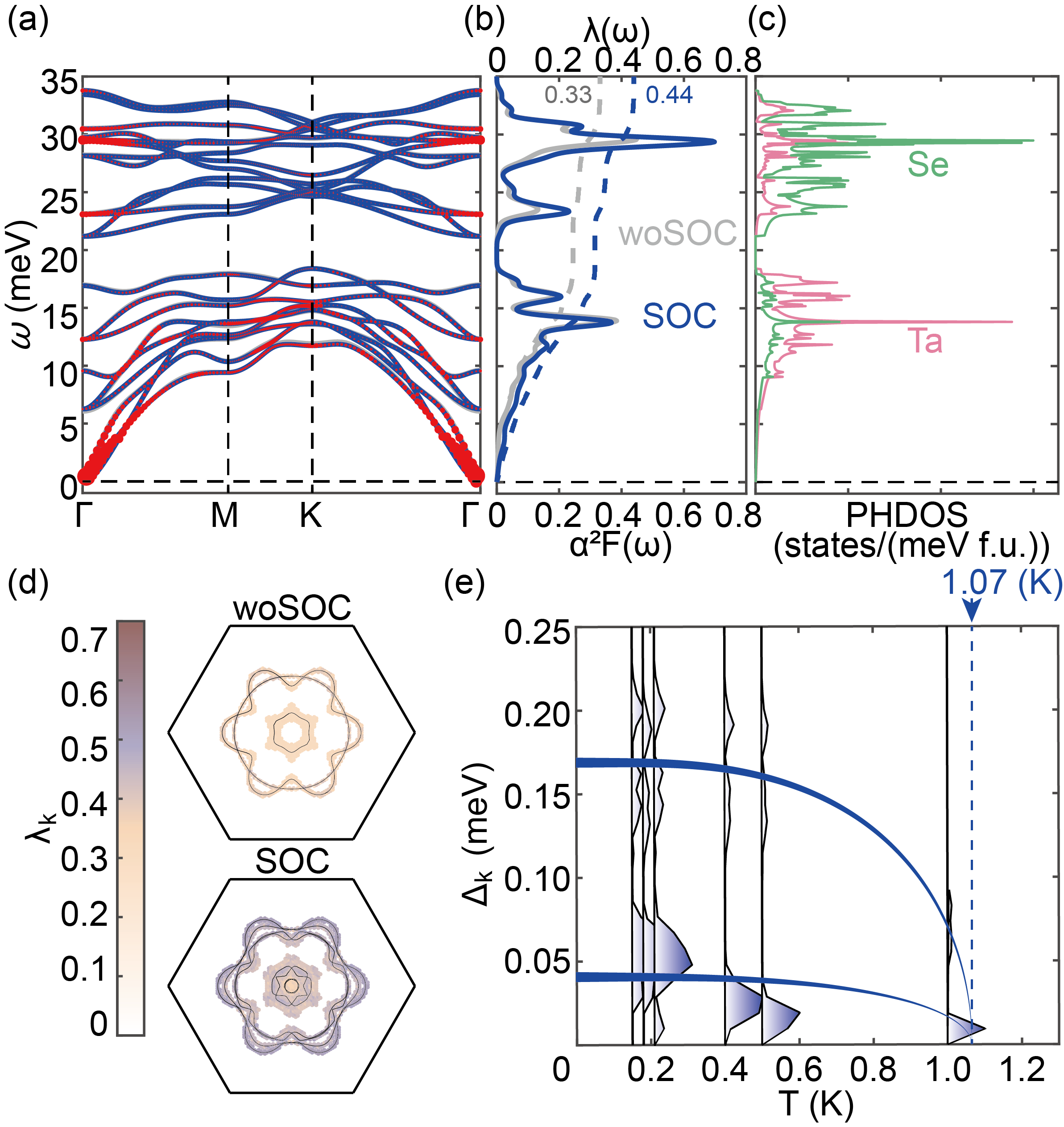}
  \caption{
  \label{fig7} 
  Calculated properties of $1H_{\mathrm{Ta}}$-Ta$_3$Se$_4$.
  (a) $\omega_{\boldsymbol{q}\nu}$ with (blue) and without (grey) SOC. 
  The size of red dots represent the value of $\lambda_{\boldsymbol{q}\nu}$ in the presence of SOC.  
  (b) $\alpha^2F(\omega)$ along with $\lambda(\omega)$, calculated with and without SOC.
  (c) Projected PHDOS onto the vibrations of Se and Ta atoms.     
  {
  (d) Distribution of $\boldsymbol{k}$-resolved EPC constants near the Fermi surface in the presence of SOC.
  Only the states within an energy window of $\pm$35 meV are shown.
  (e) Histograms of $\Delta(\boldsymbol{k},T)$ at various temperatures in the presence of SOC.
  The blue curve represents a BCS fit of the energy gap.
  }}
\end{figure}

Interestingly, in contrast to the slight increase of $N(0)$ due to SOC in $2H_{\mathrm{Ta}}$-Ta$_3$Se$_4$ and $1H_{\mathrm{hollow}}$-Ta$_3$Se$_4$, the $N(0)$ in $1H_{\mathrm{Ta}}$-Ta$_3$Se$_4$ substantially increases after involving SOC, as shown in Fig.~\ref{fig6}(b).
This primarily arises from  the increased electronic states near the $\Gamma$ point, where the contribution to $N(0)$ is significantly enhanced, particularly along $\Gamma$--M direction, as shown in Fig.~\ref{fig6}(f).
This results from SOC-induced band splitting at energies slightly above  the Fermi level near the $\Gamma$ point, as shown in Fig.~\ref{fig6}(a).
The split band is shifted downward to energy levels very close to the Fermi level, exhibiting relatively flat dispersion, which results in the emergence of a peak in the DOS at the Fermi level (Fig.~\ref{fig6}(b)).
The appearance of the DOS peak increases $N(0)$ from {3.00} /eV/f.u. without SOC to {5.02} /eV/f.u. with SOC included.

{
The SOC induced $N(0)$ increase is found to be crucial for triggering superconductivity in $1H_{\mathrm{Ta}}$-Ta$_3$Se$_4$.
A comparison of $\lambda_{\boldsymbol{k}}$ calculated with and without SOC, as depicted in Fig.~\ref{fig7}(d), shows an overall enhancement of $\lambda_{\boldsymbol{k}}$ upon incorporating SOC. 
This enhancement is physically intuitive, as the rise in $N(0)$ provides a greater number of available channels for EPC. 
This is also consistent with the computational result, where $T^{\text{aniso}}_{\mathrm{c}}$ is calculated to be higher with SOC (1.07 K) compared to without SOC (0.83~K), as shown in Fig.~\ref{fig7}(e) and Fig.~S10(b) \cite{SM}, respectively.
The calculated $\alpha^2F(\omega)$ and $\lambda$ also display consistent results.
As the calculated $\lambda$ increases from 0.33 to 0.44 after SOC is included, as shown in Fig.~\ref{fig7}(b).
This further confirms the important role of SOC plays in the superconductivity of this system.
Therefore, we have shown the SOC-enhanced superconductivity in $1H_{\mathrm{Ta}}$-Ta$_3$Se$_4$, with comparable value of $T_{\mathrm{c}}$ to TaSe$_2$ without intercalation, due to substantial increase of electronic states near the $\Gamma$ point.
}
We also note that, although the $N(0)$ in $1H_{\mathrm{Ta}}$-Ta$_3$Se$_4$ is significantly higher than that of $1H_{\mathrm{hollow}}$-Ta$_3$Se$_4$, the EPC $\lambda$  is slightly lower (0.44 vs 0.64). 
This discrepancy arises due to the significantly lower $\langle |g|^2\rangle$, which is 660 meV$^2$ in $1H_{\mathrm{Ta}}$-Ta$_3$Se$_4$ compared to that of 1850 meV$^2$ in $1H_{\mathrm{hollow}}$-Ta$_3$Se$_4$ (see Tab.~\ref{tab1}).

\section{Conclusion and discussion}

In summary, we have computationally studied the crystal structures, electronic structures, phonons, EPC, and superconductivity of three candidate crystals of Ta$_3$Se$_4$, leading to the following findings.

Firstly, we find rich charge orders in Ta$_3$Se$_4$, arising from the different interlayer stacking  between two parent TaSe$_2$ layers, and intercalation sites being occupied.
When the two parent TaSe$_2$ layers stack in $1H$ order, with the hollow sites being occupied by the intercalants, this leads to the structure of $1H_{\mathrm{hollow}}$-Ta$_3$Se$_4$, which is observed in experiment.
This structure exhibits a $\sqrt{3} \times \sqrt{3}$ CDW, coexisting with superconductivity with an estimated {$T^{\text{aniso}}_{\mathrm{c}}$ of 1.06 K}.
The calculated $\sqrt{3} \times \sqrt{3}$ CDW order is in nice agreement with experiment.
When the stacking order is $1H$, with the intercalants occupying the sites aligned with the parent Ta atoms in the out-of-plane direction, this leads to the structure of $1H_{\mathrm{Ta}}$-Ta$_3$Se$_4$.
This structure exhibits an absence of CDW, with {SOC-enhanced} superconductivity, whose {$T^{\text{aniso}}_{\mathrm{c}}$ is estimated to be 1.07 K}.
When the stacking order is $2H$, with the intercalants occupying the same sites as $1H_{\mathrm{Ta}}$-Ta$_3$Se$_4$, this leads to the structure of $2H_{\mathrm{Ta}}$-Ta$_3$Se$_4$.
This structure exhibits a $2 \times 2$ CDW, coexisting with superconductivity with an estimated {$T^{\text{aniso}}_{\mathrm{c}}$ of 1.34~K}.

Secondly, in all the three Ta$_3$Se$_4$ structures, the parent TaSe$_2$ layers consistently display suppressed CDW.
Meanwhile, the absence, or emergence of new CDW orders in the intercalation layers depends on the interlayer stacking sequences and occupied intercalation sites.
In $1H_{\mathrm{hollow}}$-Ta$_3$Se$_4$, the $T_{\mathrm{CDW}}$ is estimated to be 167 K, by considering anharmonicity and  quantum fluctuations at finite temperatures, within SSCHA approximation.
This $T_{\mathrm{CDW}}$ is higher than that of bilayer TaSe$_2$.
The  $T_{\mathrm{CDW}}$ is expected to be significantly higher in 2$H_{\mathrm{Ta}}$-Ta$_3$Se$_4$ compared to $1H_{\mathrm{hollow}}$-Ta$_3$Se$_4$ and bilayer TaSe$_2$, based on comparisons of energy gains due to CDW formation, variations in Ta-Ta distances, and phonon energies at the $\boldsymbol{q}_{\mathrm{CDW}}$ among these structures.
This suggests that the self-intercalation in  TaSe$_2$ tends to suppress CDW in parent TaSe$_2$ layers.
Furthermore, the CDWs induced by the self-intercalation in the intercalation layers are  enhanced compared to parent TaSe$_2$.

\begin{table*}[t]
  \caption{
  \label{tab1}
  The parameters of each structure, where the N(0) is given in units of /eV/c.f., the $T_\mathrm{c}$ and $\omega_{\log}$ is given in units of K.
  The averaged EPC matrix element square, $\langle|g|^2\rangle = 1/N(0)\int \alpha^2F(\omega) \mathrm{d}\omega$, is in the unit of $10^3$ $\mathrm{meV}^2$.
  The data in parentheses are calculated without SOC.
  }
  \begin{ruledtabular}
  \begin{tabular}{cccccccccc}
  \multicolumn{2}{c}{System}&CDW& $N(0)$ & $\langle|g|^2\rangle$  & $\omega_{\log}$  &$\lambda$&{$T^{\text{ME}}_{\mathrm{c}}$}&{$T^{\text{aniso}}_\mathrm{c}$}&Exp. $T_\mathrm{c}$\\
  \hline
  \multirow{3}*{\centering 2\textit{H}-$\mathrm{TaSe_2}$}
  &Bulk\footnote{The experimental data are from \cite{wilson1975charge,bhoi2016interplay,freitas2016strong}, and the computational data are from \cite{lian2019coexistence}.}
  &3 $\times$ 3&2.8&  &202.73&0.4&0.14&& 0.1--0.15\\
  &multilayer\footnote{The experimental data are from \cite{Wu_2019,galvis2013scanning}.}
  &&&&&&&&1-1.4\\
  &monolayer\footnote{The computational data are from \cite{lian2019coexistence,lian2023interplay}.}
  &3 $\times$ 3&2.46& &79.07&0.71&1.8&\\
  \hline
  \multirow{3}*{$\mathrm{Ta_3Se_4}$\footnote{This work.}}& $1H_{\mathrm{hollow}}$ &$\sqrt{3} \times \sqrt{3}$&1.98 (1.65)& 1.85 (2.82)  &110.92 (84.64)&0.64 (0.92)&1.79 (3.82)&{1.06 (7.79)}\\
                                  & $2H_{\mathrm{Ta}}$  &2 $\times$ 2&2.46 (2.17)& 1.24 (1.45)  & 109.06 (121.13)&0.48 (0.47)&0.39 (0.42)&{(1.34)}\\
                                  & $1H_{\mathrm{Ta}}$  &none&{5.02} ({3.00})& {0.66} ({0.87}) &{134.68} ({150.44})&{0.44} (0.33)&{0.27} (0.01)&{1.07 (0.83)}\\
  \end{tabular}
  \end{ruledtabular}
\end{table*}

Thirdly, SOC tends to increase $N(0)$ while simultaneously suppressing EPC matrix elements in Ta$_3$Se$_4$ by comparing these quantities across different structures, as tabulated in Tab.~\ref{tab1}.
The interplay between these two competing SOC-induced effects varies among the three Ta$_3$Se$_4$ crystal structures, resulting in distinct impacts on their respective $T^{\text{ME}}_{\mathrm{c}}$ and $T^{\text{aniso}}_{\mathrm{c}}$.
In $1H_{\mathrm{hollow}}$-Ta$_3$Se$_4$, the reduction in EPC matrix elements has a more pronounced impact than the enhancement of $N(0)$, ultimately resulting in a decrease in $\lambda$ and two types of $T_{\mathrm{c}}$.
In contrast, for $1H_{\mathrm{Ta}}$-Ta$_3$Se$_4$, the $N(0)$ is significantly enhanced due to SOC-induced band splitting.
This enhancement results in a substantial increase in $N(0)$ near the $\Gamma$ point, {which provides more available channels for EPC, giving rise to increases in both $T^{\text{ME}}_{\mathrm{c}}$ and $T^{\text{aniso}}_{\mathrm{c}}$.}
The case is different for $2H_{\mathrm{Ta}}$-Ta$_3$Se$_4$, where the influence of SOC on $N(0)$ and EPC matrix elements is  minor and tends to cancel out.

Finally, the CDW in $2H_{\mathrm{Ta}}$-Ta$_3$Se$_4$ is found to enhance $N(0)$ relative to its non-CDW phase, primarily attributed to the formation of energy gaps in the occupied states.
This unusual behavior distinguishes itself from a typical metal-to-insulator Peierls transition, where energy gaps generally form at the Fermi level.

Our work unveils the properties in self-intercalated bilayer TaSe$_2$,  with a focus on the CDW, superconductivity, magnetism, and proposes candidates  for the study of the interplay between these orders.

\begin{acknowledgements} 
    This work is supported by National Natural Science Foundation of China 11804118, Guangdong Basic and Applied Basic Research Foundation (Grant No.{2025A1515010219}, 2021A1515010041), and the Science and Technology Planning Project of Guangzhou (Grant No. 202201010222). 
    The Calculations were performed on  high-performance computation cluster of Jinan University, and Tianhe Supercomputer System.
\end{acknowledgements}

{All data available from the authors upon reasonable request.}

\bibliographystyle{apsrev4-2}
%
\end{document}